\documentclass[12pt]{iopart}
\usepackage{graphicx}
\begin{document}
\title{New flow observables}

\author{R S Bhalerao$^1$, M Luzum$^2$ and J-Y Ollitrault$^2$}

\address{$^1$ Department of Theoretical Physics, TIFR,
   Homi Bhabha Road, Colaba, Mumbai 400 005, India}
\address{$^2$ Institut de physique th\'eorique, CNRS/URA2306, CEA Saclay, 91191 Gif-sur-Yvette, France}

\ead{jean-yves.ollitrault@cea.fr}

\begin{abstract}
Event-by-event fluctuations of the initial transverse density profile
result in a collective flow pattern which also fluctuates event by
event. We propose a number of new correlation observables to
characterize these fluctuations and discuss how they should be
analyzed experimentally. 
We argue that most of these quantities can be measured at RHIC and LHC. 
\end{abstract}    


Thermalization of the matter produced in ultrarelativistic
nucleus-nucleus collisions results in strong collective motion. 
The clearest experimental signature of such collective flow is obtained
from azimuthal correlations between outgoing particles. 
There is now a growing consensus that correlations between particles
emitted at large relative pseudorapidity $\Delta\eta$ are all due to 
flow~\cite{Luzum:2010sp}.  
Heavy-ion experiments at the CERN Large Hadron Collider (LHC) will be
able to carry out correlation analyses with unprecedented accuracy due
to large multiplicities, and the large detector coverage. 
We propose a number of new independent flow measurements and discuss their
feasibility. 

The flow hypothesis is that particles in a given event are emitted
{\it independently\/} according to some azimuthal distribution.  The
most general distribution can be written as a sum of Fourier components, 
\begin{equation}
\label{Fourier}
\frac{dN}{d\varphi}=\frac{N}{2\pi}\left(1+2\sum_{n=1}^{\infty}
    v_n\cos(n\varphi-n\Psi_n)\right),
\end{equation}
where $v_n$ is the $n^{th}$ flow harmonic~\cite{Voloshin:1994mz} and
$\Psi_n$ the corresponding reference angle, all of which fluctuate event-by-event. 

In practice, 
one cannot exactly reconstruct the underlying 
probability distribution from the finite sample of particles
emitted in a given event. 
All known information about $v_n$ is inferred from azimuthal
correlations. 
Generally, a $k$-particle correlation is of the type
\begin{equation}
\label{defcor}
v\{n_1,n_2,\ldots,n_k\}=\left\langle \cos\left(n_1\varphi_1+\ldots+n_k\varphi_k\right)\right\rangle,
\end{equation}
where $n_1,\ldots,n_k$ are integers, $\varphi_1,\ldots,\varphi_k$ are
azimuthal angles of particles belonging to the same event, 
and angular brackets denote average over multiplets of particles and
events in a centrality class. Since the impact parameter orientation is uncontrolled, the only
measurable correlations have azimuthal symmetry: $n_1+\ldots+n_k=0$.  

Inserting Eq.~(\ref{Fourier}) into Eq.~(\ref{defcor}) gives 
\begin{equation}
\label{corflow}
v\{n_1,\ldots,n_k\}=\left\langle v_{n_1}\ldots v_{n_k}\cos(n_1\Psi_{n_1}+\ldots+n_k\Psi_{n_k})\right\rangle,
\end{equation}
where the average is now only over events.
To the extent that correlations are induced by collective flow, 
azimuthal correlations measure moments of the flow distribution.

In practice, the average over particles in Eq.~(\ref{defcor}) is a
weighted average: in a given harmonic $n$, one gives more weight to
particles which have larger $v_n$ in order to increase the 
resolution. 
Our goal here is to characterize initial-state fluctuations,
which are approximately independent of
rapidity~\cite{Dumitru:2008wn}. 
Weights should therefore be chosen independent of (pseudo)rapidity,  
a nonstandard choice for odd harmonics~\cite{Poskanzer:1998yz}. 

The simplest $v_n$ measurement is the pair 
correlation~\cite{Wang:1991qh}, which corresponds to  the event-averaged root-mean-square $v_n$
\begin{equation}
\label{vn2}
v_n\{2\}\equiv \sqrt{v\{n,-n\}}\simeq \sqrt{\langle v_n^{\ 2} \rangle}.
\end{equation}
At this Conference, measurements of $v\{n,-n\}$ were presented for
$n=2,\cdots,6$, but $v\{1,-1\}$ has not yet been directly analyzed
with rapidity-independent weights~\cite{Luzum:2010fb}.  

Higher-order correlations yield higher moments of the $v_n$ distribution:
\begin{equation}
\label{vn4}
v\{n,n,-n,-n\}\equiv2v_n\{2\}^4-
v_n\{4\}^4\simeq\langle v_n^{\ 4} \rangle,
\end{equation}
where we have used the standard notation  $v_n\{4\}$ for the 4-particle cumulant~\cite{Borghini:2001vi}.
Finally, one can construct correlations involving mixed harmonics. 
The first non-trivial correlations between $v_1$, $v_2$ and $v_3$
are~\cite{Bhalerao:2011yg}  
\begin{eqnarray}
\label{corr123}
&v_{12}\equiv v\{1,1,-2\},\ \ \ \ \ \ \ \ \ 
&v_{13\phantom{2}}\equiv v\{1,1,1,-3\}, \nonumber \\
&v_{23}\equiv v\{2,2,2,-3,-3\},\ \ \ \ \ \ \ \ \ 
&v_{123}\equiv v\{1,2,-3\}.
\end{eqnarray}
\begin{figure}
\includegraphics[width=18pc]{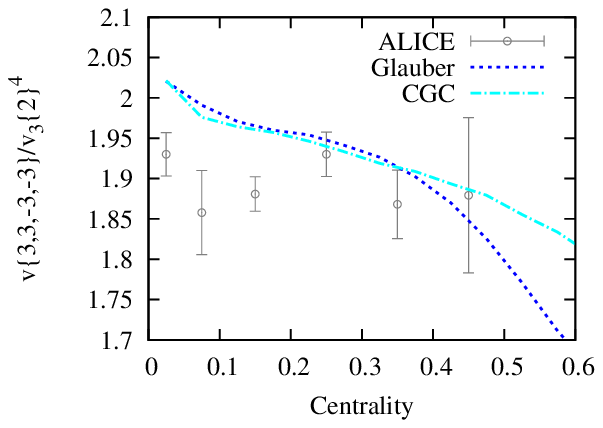}
\includegraphics[width=18pc]{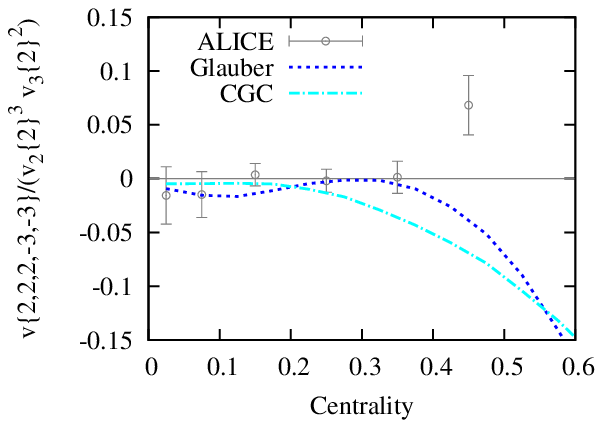}
\caption{\label{alicedata}
Curves are theoretical predictions for Pb-Pb collisions at
2.76~TeV~\cite{Bhalerao:2011yg} using the Monte-Carlo Glauber
model~\cite{Alver:2008aq} (dotted lines) and the Monte-Carlo KLN
model~\cite{Drescher:2007ax,ALbacete:2010ad} (dashed lines). Data
points are from ALICE~\cite{Collaboration:2011vk}.
}
\end{figure}
These mixed correlations involve angular correlations between
$\Psi_1$, $\Psi_2$ and $\Psi_3$. 
In order to single out these angular correlations, it is natural to
scale $v\{n_1,\ldots,n_k\}$ by $v_{n_1}\{2\}\cdots v_{n_k}\{2\}$ in
order to obtain a number of order unity. 
Figure~\ref{alicedata} displays predictions for two such 
ratios~\cite{Bhalerao:2011yg} together with data recently released by
the ALICE collaboration~\cite{Collaboration:2011vk}. 
The measured $\langle v_3^4\rangle/\langle v_3^2\rangle^2$
(Figure~\ref{alicedata}, left) is  
compatible with theoretical prediction except for $0\%-20\%$ central
collisions, where the prediction is close to 2 (corresponding to
Gaussian fluctuations~\cite{Voloshin:2007pc}) while the measured value
is close to 1.9. 
The second ratio (Figure~\ref{alicedata}, right) 
measures the correlation between the orientations of the ellipse
($\Psi_2$) and the triangle ($\Psi_3$). It is
predicted to be small up to 50\% centrality, in qualitative agreement
with the measurement. Note, however, that neither model agrees 
quantitatively with data. 

All correlations should be analyzed in such a way as to isolate 
the correlation induced by collective flow from other ``nonflow''
effects. 
An important nonflow effect is global momentum conservation. 
It contributes to $v\{1,-1\}$~\cite{Borghini:2000cm} and
also, to a lesser extent, to $v\{1,1,-2\}$~\cite{Borghini:2003ur}. 
This nonflow effect can be suppressed~\cite{Luzum:2010fb} simply by using
the weight $w=p_t-\langle p_t^2\rangle/\langle
p_t\rangle$ 
for at least one of the particles with Fourier harmonic 1. 

Other nonflow effects are correlations between a small number of
particles ---typically pairs of particles--- and they are suppressed by
putting rapidity gaps between particles~\cite{Luzum:2010sp}. 
In order to determine where rapidity gaps are important, we estimate 
nonflow effects by assuming that particles are emitted in collinear
pairs. If $M$ particles are observed in each event, the probability
that two random particles belong to the same pair is $1/(M-1)\simeq 1/M$. 
Consider the first correlation in  Eq.~(\ref{corr123}), $v_{12}$,
which involves three particles. There are different nonflow contributions
to this correlation corresponding to the different
pairings. Pairing the first two particles (with harmonic 1) gives a nonflow correlation
of order $(v_2)^2/M$, while pairing 1 or 2 with 3 gives a correlation of
order $(v_1)^2/M$. 
Since $v_2\gg v_1$, it is important to put a rapidity gap between the
first two particles. On the other hand, there is no restriction for
the third particle. 
For $v_{13}$, a similar discussion shows that there must be rapidity
gaps between the first three particles (again those with harmonic 1). 
For $v_{23}$, nonflow effects are small and rapidity gaps are not
required. 
Finally, for $v_{123}$, the largest nonflow correlation is between the
first and the third particle (harmonics 1 and 3) and is of order
$v_2^2/M$. The next-to-largest is between the first two 
particles, of order $v_3^2/M$, which is much 
smaller, except for central collisions.
\begin{table}
\caption{\label{stat}Number of events needed in 
  the $0\%-5\%$ centrality class in order to measure various
  quantities within $5\%$.}
\begin{indented}
\item[]
\begin{tabular}{lr|lll}
\br
quantity &S& ALICE&ATLAS or CMS& STAR or PHENIX\\
\mr
$v\{2,-2\}$&2& $3\times 10^2$ &  $2\times 10^2$ &  $9\times 10^2$ \\
$v\{3,-3\}$&2& $6\times 10^2$ &  $2\times 10^2$ &  $2\times 10^3$ \\
$v\{1,-1\}$&2& $1\times 10^4$ &  $2\times 10^3$ &  $5\times 10^4$ \\
$v\{1,2,-3\}/(v_1v_2v_3)$ &1&$9\times 10^3$ &  $2\times 10^3$ &  $5\times 10^4$ \\
$v\{2,2,-2,-2\}/v_2^4$&8& $2\times 10^4$ &
$4\times 10^3$ &  $1\times 10^5$ \\

$v\{3,3,-3,-3\}/v_3^4$ &8&$5\times 10^4$ &
$7\times 10^3$ &  $4\times 10^5$ \\
$v\{1,1,-2\}/(v_1^2v_2)$&2& $8\times 10^4$ &  $8\times 10^3$ &  $5\times 10^5$ \\
$v\{2,2,2,-3,-3\}/(v_2^3v_3^2)$&12& $9\times 10^4$ &
$1\times 10^4$ &  $1\times 10^6$ \\
$v\{1,1,1,-3\}/(v_1^3v_3)$&6& $3\times 10^6$ &
$1\times 10^5$ &  $5\times 10^7$ \\
$v\{1,1,-1,-1\}/ v_1^4$&8& $2\times 10^7$ &
$4\times 10^5$ &  $3\times 10^8$ \\
\br
\end{tabular}
\end{indented}
\end{table}

The limiting factor in the ability to measure these high-order
correlations is statistics.  
The statistical error is 
\begin{equation}
\frac{\delta v\{n_1,\cdots,n_k\}}{v_{n_1}\{2\}\cdots v_{n_k}\{2\}}=
\sqrt{\frac{S}{2N_{\rm evts}}\left(1+\frac{1}{\chi_{n_1}^2}\right)\cdots\left(1+\frac{1}{\chi_{n_k}^2}\right)}, 
\end{equation}
where $\chi_n\equiv v_n\{2\}\sqrt{M}$ is the resolution 
parameter~\cite{Ollitrault:1997di} in harmonic $n$, and $S$ is a
symmetry factor which is given in Table~\ref{stat} for the various
correlations. 
Table~\ref{stat} lists the number of events required in each
experiment to measure the various ratios for central collisions. We
have assumed $v_1\{2\}=0.74\%$, $v_2\{2\}=2.40\%$, $v_3\{2\}=1.90\%$
and $M=2000$, $6000$ and $900$ respectively for ALICE, CMS/ATLAS
and STAR/PHENIX. 

We have introduced new flow observables, most of which can be measured
accurately at RHIC and LHC with 
a modest number of events. The most challenging measurements are
$v\{1,1,1,-3\}$ and $v\{1,1,-1,-1\}$, which require wide 
pseudorapidity coverage or large statistics. 
These new observables will constrain models of initial-state 
fluctuations and deepen our knowledge of the collision dynamics.

\ack
We thank R. Snellings for providing the experimental ALICE data. 
This work is funded by ``Agence Nationale de la Recherche'' under grant
ANR-08-BLAN-0093-01 and by CEFIPRA under project 4404-2. 

\section*{References}
\bibliography{fluctuations}

\providecommand{\newblock}{}
\begin{thebibliography}{10}
\expandafter\ifx\csname url\endcsname\relax
  \def\url#1{{\tt #1}}\fi
\expandafter\ifx\csname urlprefix\endcsname\relax\def\urlprefix{URL }\fi
\providecommand{\eprint}[2][]{\url{#2}}

\bibitem{Luzum:2010sp}
Luzum M 2011 {\em Phys.Lett.\/} {\bf B696} 499--504 (\textit{Preprint}
  \eprint{1011.5773})

\bibitem{Voloshin:1994mz}
Voloshin S and Zhang Y 1996 {\em Z.Phys.\/} {\bf C70} 665--672
  (\textit{Preprint} \eprint{hep-ph/9407282})

\bibitem{Dumitru:2008wn}
Dumitru A, Gelis F, McLerran L and Venugopalan R 2008 {\em Nucl. Phys.\/} {\bf
  A810} 91 (\textit{Preprint} \eprint{0804.3858})

\bibitem{Poskanzer:1998yz}
Poskanzer A~M and Voloshin S 1998 {\em Phys.Rev.\/} {\bf C58} 1671--1678
  (\textit{Preprint} \eprint{nucl-ex/9805001})

\bibitem{Wang:1991qh}
Wang S, Jiang Y, Liu Y, Keane D, Beavis D {\em et~al.\/} 1991 {\em Phys.Rev.\/}
  {\bf C44} 1091--1095

\bibitem{Luzum:2010fb}
Luzum M and Ollitrault J~Y 2011 {\em Phys.Rev.Lett.\/} {\bf 106} 102301
  (\textit{Preprint} \eprint{1011.6361})

\bibitem{Borghini:2001vi}
Borghini N, Dinh P~M and Ollitrault J~Y 2001 {\em Phys.Rev.\/} {\bf C64} 054901
  (\textit{Preprint} \eprint{nucl-th/0105040})

\bibitem{Bhalerao:2011yg}
Bhalerao R~S, Luzum M and Ollitrault J~Y 2011  (\textit{Preprint}
  \eprint{1104.4740})

\bibitem{Alver:2008aq}
Alver B, Baker M, Loizides C and Steinberg P 2008  (\textit{Preprint}
  \eprint{0805.4411})

\bibitem{Drescher:2007ax}
Drescher H~J and Nara Y 2007 {\em Phys.Rev.\/} {\bf C76} 041903
  (\textit{Preprint} \eprint{0707.0249})

\bibitem{ALbacete:2010ad}
Albacete J~L and Dumitru A 2010  (\textit{Preprint} \eprint{1011.5161})

\bibitem{Collaboration:2011vk}
Aamodt K {\em et~al.\/} (ALICE Collaboration) 2011  (\textit{Preprint}
  \eprint{1105.3865})

\bibitem{Voloshin:2007pc}
Voloshin S~A, Poskanzer A~M, Tang A and Wang G 2008 {\em Phys. Lett.\/} {\bf
  B659} 537--541 (\textit{Preprint} \eprint{0708.0800})

\bibitem{Borghini:2000cm}
Borghini N, Dinh P~M and Ollitrault J~Y 2000 {\em Phys. Rev.\/} {\bf C62}
  034902 (\textit{Preprint} \eprint{nucl-th/0004026})

\bibitem{Borghini:2003ur}
Borghini N 2003 {\em Eur.Phys.J.\/} {\bf C30} 381--385 (\textit{Preprint}
  \eprint{hep-ph/0302139})

\bibitem{Ollitrault:1997di}
Ollitrault J~Y 1997  (\textit{Preprint} \eprint{nucl-ex/9711003})

\end{thebibliography}

\end{document}